\documentclass[sigconf,nonacm=true]{acmart}
\AtBeginDocument{%
  }

\usepackage{amsmath}

\usepackage{amssymb}

\begin{document}

\title{Can Online GenAI Discussion Serve as Bellwether for Labor Market Shifts?}

\author{Shurui Cao}
\email{shuruic@andrew.cmu.edu}
\affiliation{
  \institution{Heinz College, Carnegie Mellon University}
  \city{Pittsburgh}
  \state{PA}
  \country{USA}
}

\author{Wenyue Hua}
\affiliation{%
  \institution{University of California, Santa Barbara}
  \city{Santa Barbara}
  \state{CA}
  \country{USA}}
\email{wenyuehua@microsoft.com}

\author{William Yang Wang}
\affiliation{%
  \institution{University of California, Santa Barbara}
  \city{Santa Barbara}
  \state{CA}
  \country{USA}}
\email{william@cs.ucsb.edu}

\author{Hong Shen}
\affiliation{%
  \institution{Carnegie Mellon University}
  \city{Pittsburgh}
  \state{PA}
  \country{USA}
}
\email{hongs@andrew.cmu.edu}

\author{Fei Fang}
\affiliation{%
  \institution{Carnegie Mellon University}
  \city{Pittsburgh}
  \state{PA}
  \country{USA}
}
\email{feifang@cmu.edu}

\newcommand{\blue}[1]{\textcolor{blue}{#1}}

\renewcommand{\shortauthors}{Trovato et al.}

\begin{abstract}
The rapid advancement of Large Language Models (LLMs) has generated considerable speculation regarding their transformative potential for labor markets. However, existing approaches to measuring AI exposure in the workforce predominantly rely on concurrent market conditions, offering limited predictive capacity for anticipating future disruptions. This paper presents a predictive study examining whether online discussions about LLMs can function as early indicators of labor market shifts. We employ four distinct analytical approaches to identify the domains and timeframes in which public discourse serves as a leading signal for employment changes, thereby demonstrating its predictive validity for labor market dynamics. Drawing on a comprehensive dataset that integrates the REALM corpus of LLM discussions, LinkedIn job postings, Indeed employment indices, and over 4 million LinkedIn user profiles, we analyze the relationship between discussion intensity across news media and Reddit forums and subsequent variations in job posting volumes, occupational net change ratios, job tenure patterns, unemployment duration, and transitions to GenAI-related roles across thirteen occupational categories. Our findings reveal that discussion intensity predicts employment changes 1-7 months in advance across multiple indicators, including job postings, net hiring rates, tenure patterns, and unemployment duration. These findings suggest that monitoring online discourse can provide actionable intelligence for workers making reskilling decisions and organizations anticipating skill requirements, offering a real-time complement to traditional labor statistics in navigating technological disruption.
\end{abstract}




\received{20 February 2007}
\received[revised]{12 March 2009}
\received[accepted]{5 June 2009}

\maketitle

\section{Introduction}
The rapid advancement of Large Language Models (LLMs)~\cite{zhao2023survey, kasneci2023chatgpt} has sparked intense speculation about their transformative impact on labor markets~\cite{eloundou2023gpts, humlum2025large}. Since the public release of ChatGPT in November 2022, discussions about AI's potential to automate jobs have moved from academic circles to mainstream discourse. Previous research has shown measurable changes in employment patterns~\cite{hatzius2023potentially, demirci2025ai, lin2024tech}, including estimates that about 19\% of American workers are in occupations with AI exposure~\cite{kochhar2023which} and evidence of a 21\% decline in demand for automation-prone roles post-ChatGPT~\cite{demirci2025ai}. However, these studies typically focus on the existent labor market and offer limited predictive value for workers and organizations navigating this transition.


This paper addresses a critical gap in the literature by investigating predictive changes in the job market relevant to AI's impact. Unlike traditional approaches that measure current AI exposure across pre-defined occupational classifications such as O*NET~\cite{cifuentes2010use, peterson2001understanding, handel2016net}, we leverage public discourse~\cite{Adams2023MoreThanWords, Huynh2023MuskBitcoin, Buz2024Reddit, Kokil2019election, Jinping2924election, Qin2929COVID, Wang2023disease, Higgins2020COVID} as a leading indicator of employment changes, representing the most commonly available and observable information about LLMs. We recognize that public awareness and discussion of technological capabilities often precede their implementation in workplaces, potentially providing valuable signals for labor market transitions. Therefore, we analyze the relationship between LLM discussion intensity across news media and Reddit forums and subsequent changes in multiple labor market dimensions. To study this question, we create a comprehensive dataset combining REALM~\cite{cheng2025realm} -- a dataset investigating online discussion on LLM, LinkedIn job postings~\cite{li2020deep, mccabe2017social}, Indeed Job Postings Index~\cite{ollech2021seasonal}\footnote{\text{https://github.com/hiring-lab/job\_postings\_tracker}}, and over 4 million LinkedIn user profiles. Based on this dataset, we examine how variations in public discourse relate to and potentially predict changes in job posting volumes, occupational net change ratios, job tenure patterns, unemployment duration, and transitions to GenAI-related roles across thirteen occupational categories. Therefore, in this paper, we focus on two research questions: 

\begin{itemize}
\item RQ1: How does GenAI adoption affect employment dynamics in terms of job tenure patterns, unemployment duration, hiring flows, and career transitions between GenAI and non-GenAI roles?
\item RQ2: Can online discussion intensity about LLMs serve as a leading indicator for changes in job posting volumes, net hiring ratios, tenure stability, and unemployment patterns across occupations?
\end{itemize}

Based on our analysis, we find that: (1) GenAI workers experience distinct career patterns, shifting toward medium-tenure positions (4-12 months) while showing reduced long-term retention (12+ months) compared to non-GenAI workers. (2) Online LLM discussion can serve as a leading indicator for labor market changes, with discussion spikes on Reddit and News platforms preceding employment shifts by 1-7 months across occupations. Specifically, knowledge-intensive occupations like Computer \& Math, Arts, and Education demonstrate the strongest predictive signals between online discourse and subsequent changes in job postings, hiring flows, tenure, and unemployment duration.

The findings have significant implications for labor market participants and policymakers. If online discourse can reliably signal impending employment shifts, it would provide workers with actionable intelligence for reskilling decisions and career planning, while enabling organizations to anticipate changing skill requirements. By establishing the predictive validity of digital discourse as a labor market indicator, this research contributes to both the empirical understanding of AI's economic impact and the development of practical tools for navigating technological disruption.

\begin{figure}[htbp]
    \centering
    \includegraphics[width=1 \linewidth]{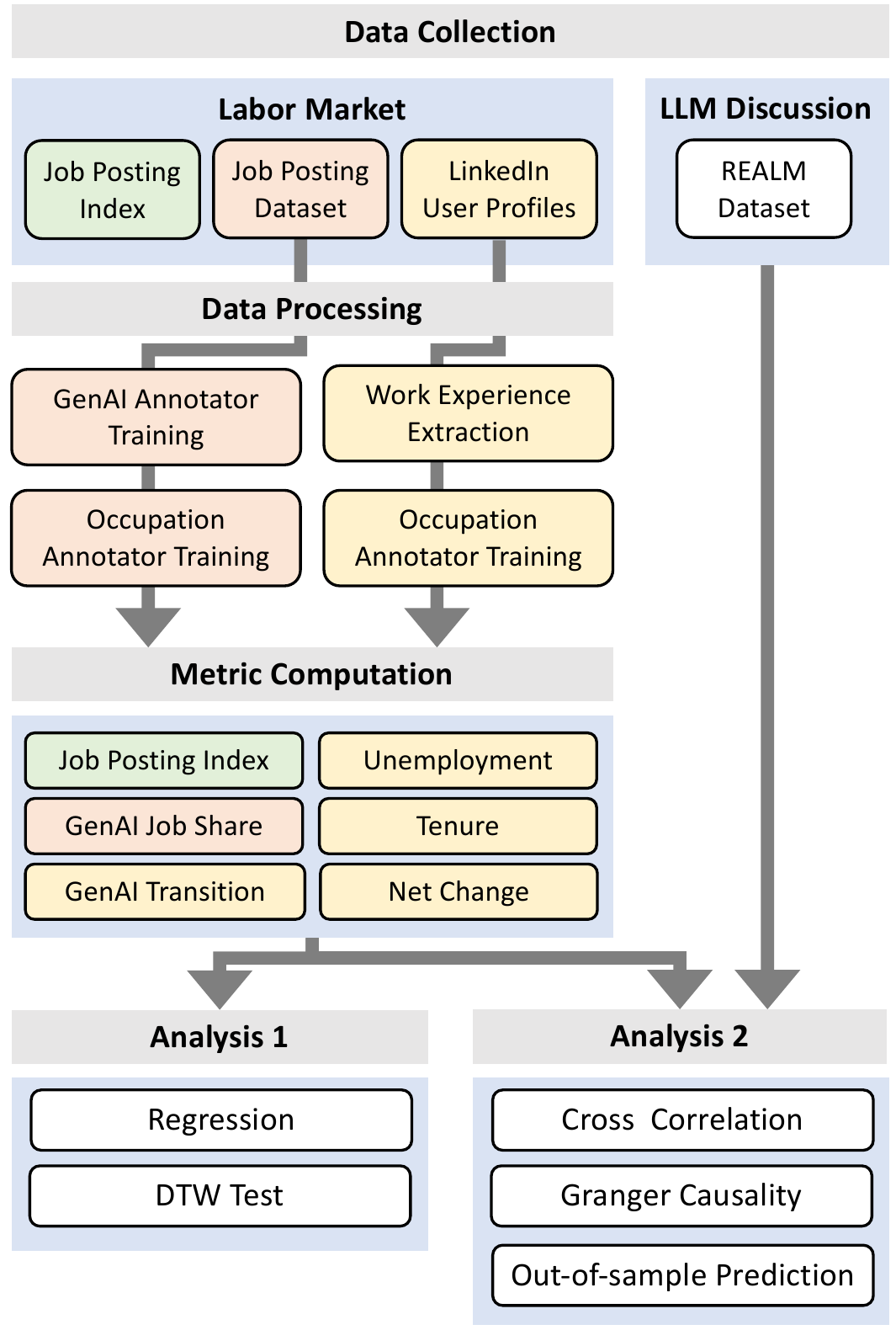}
    \caption{Overview of the analysis pipeline. We compile data from online LLM discussions, job postings, and LinkedIn user profiles, apply systematic filtering, occupation classification, and GenAI labeling, and then conduct analysis for the two research questions.}
    \label{fig:pipeline}
\end{figure}


The remainder of this paper is organized as follows. Section 2 reviews related work on GenAI and its impact on the labor market. Section 3 describes our data sources and data processing methods. Section 4 details the construction of our labor market metrics: net change ratios, tenure distributions, unemployment duration, and GenAI transition rates. Section 5 and Section 6 focus on the two research questions, respectively. Section 7 concludes the paper.

\section{Related Works}

\paragraph{\textbf{Generative AI Capabilities and Applications}}
Generative AI has demonstrated transformative capabilities across diverse professional domains, fundamentally reshaping how work is conducted in the modern labor market. In creative fields such as Art \& Media~\cite{vyas2022ethical, el2024generative}, GenAI tools enable rapid content generation, from graphic design~\cite{dehman2023graphic,fleischmann2024generative} to video editing~\cite{zhang2022ai, othman2023ai}, while in knowledge-intensive sectors like Legal and Business \& Finance~\cite{lee2024comprehensive}, these systems excel at document analysis~\cite{Ma2025Long}, contract review~\cite{williams2024generative, khalid2025evaluating}, and financial modeling~\cite{joshi2025advancing}. Technical domains, including Architecture \& Engineering and Computer \& Math have embraced GenAI for code generation~\cite{idrisov2024program}, CAD design optimization~\cite{li2025generative, zhu2024designing}, and complex problem-solving, whereas Life \& Science professionals leverage these tools for drug discovery~\cite{bian2021generative, tang2021generative}, research synthesis~\cite{clark2025generative, khosravi2025exploring}, and experimental design. Healthcare~\cite{zhang2023generative, varghese2024chatgpt} sectors apply these tools for diagnostic assistance~\cite{abdullahi2024learning, horak2023healthcare}, treatment planning~\cite{grupac2023generative, atkinson2023generative}, and patient communication~\cite{traylor2025beyond}. This broad adoption across different occupations shows GenAI's versatility in augmenting human capabilities, automating routine tasks, and enabling professionals to focus on higher-value activities, thereby catalyzing significant shifts in labor market dynamics and worker productivity~\cite{salari2025impacts}.

\paragraph{\textbf{Generative AI and Labor Market Impacts}}
A growing body of research studies how advances in GenAI and LLMs affect labor demand and supply, and how these effects evolve over time. 
Several studies measure changes in job postings and task content to infer demand shifts. For example,\citet{Hampole2025Labor} analyze trends in online job postings over a decade and find that higher average exposure of tasks to AI is associated with lower labor demand. Other works compare job postings in the pre- and post-ChatGPT periods~\cite{Demirci2024FreelancingAI, Liu2023GenerateFutureWork} and find a decline in job demand for automation-prone jobs related to writing and coding compared to jobs relying on manual skills or not directly exposed to ChatGPT's core functionalities.

Prior research also measures how AI capabilities align with the tasks and skills used across occupations. Approaches include mapping AI capabilities to occupational abilities documented in the O*NET database~\cite{felten2023languagemodelerslikechatgpt, Felten2023OccupationalExposure, Zarifhonarvar2024EconomicsChatGPT}, using structured rubrics to rate alignment between LLM skills and tasks~\cite{eloundou2023gptsgptsearlylook}, and collecting domain expert judgments~\cite{shao2025futureworkaiagents}.
These studies consistently show within-occupation heterogeneity. At the aggregate level, a substantial fraction of occupations appears impacted by LLMs, with the most exposure often concentrated in highly educated, higher-paid, white-collar roles. At the same time, exposure spans industries and wage levels, and higher-wage jobs may face larger augmentation.

\paragraph{\textbf{Online Discussions as Leading Indicators}}
Despite advances in measuring AI's labor market impacts, existing approaches face temporal limitations: job postings and surveys capture changes only after they materialize. Online platforms, where workers actively discuss emerging technologies, may offer earlier signals of these transformations. Online platforms provide real-time and high–frequency signals that could reflect public opinions and sentiment.
This makes it a useful source for tracking societal change. Researchers have used signals from social media such as Twitter and Reddit, as well as online platforms like Google Trends, to study whether these signals can indicate future changes in society, such as stock prices and financial markets~\cite{Adams2023MoreThanWords, Huynh2023MuskBitcoin, Buz2024Reddit}, election outcomes~\cite{Kokil2019election, Jinping2924election}, and disease outbreaks~\cite{Qin2929COVID, Wang2023disease, Higgins2020COVID}. 

In particular, there has been work that studies online discussion as \emph{early labor market indicators}. Studies such as~\cite{Adu2023VAR, Borup2022GoogleUnemployment, Simionescu2022GoogleUnemployment} nowcast unemployment rate using Google Trends and find that they often show strong contemporaneous correlation with official labor statistics.
Beyond search engines, researchers are increasingly mining social media, such as Twitter, and news content for labor market insights. Textual sentiment and topic frequencies in online discussions have proven valuable for economic nowcasting, such as the trends of job search and loss~\cite{Antenucci2914TwitterJobLoss, Aizenman2023JobLossTwitter, Proserpio2016TwitterJobLoss}.
While these works primarily focus on unemployment-related outcomes, our study extends the analysis to a broader set of labor market indicators.

This literature provides context for our study to connect signals from online discussions to the impacts of GenAI on the labor market. If the intensity of the online discussion on LLMs captures the workers' first-hand experiences, it could potentially move ahead of measured outcomes. 

\section{Data Sources and Preprocessing}

In this section, we describe the datasets we use for analysis and outline the pre-processing pipeline, including systematic filtering, occupation classification, and GenAI-related labeling methods used to align them for analysis.

\subsection{Data Collection}
Our analysis draws on three complementary data sources, each capturing a distinct dimension of the relationship between LLMs and labor market dynamics.

\paragraph{\textbf{LLM Discussion Data}}
We use the REALM dataset~\cite{cheng-etal-2025-realm}, which aggregates real-world LLM use cases from Reddit posts and news articles between 2020 and 2024. Each piece of discussion is annotated with the occupation categories of the end users based on O*NET classifications. This dataset provides a proxy for the \emph{perceived exposure} and \emph{potential impact} of LLMs across occupations.

\paragraph{\textbf{Job Posting Data}}
We combine two complementary sources. First, we collected approximately 254,000 U.S. job postings between June 2021 and May 2024 through Bright Data\footnote{\url{https://brightdata.com/}}, with information on job title, description, company name, and posting date. Second, we incorporate the Indeed Job Posting Index\footnote{\url{https://github.com/hiring-lab/job_postings_tracker}}, which reports seasonally adjusted percentage changes in postings (seven-day trailing average) relative to February 1, 2020. Together, these sources provide both micro-level posting information and aggregate measures of overall employer demand.

\paragraph{\textbf{LinkedIn Profile Data}}
We obtained around 4 million LinkedIn user profiles via Bright Data. To ensure data quality, we retain only users with at most ten work experiences and at most five education entries, requiring that all work experiences include company names, job titles, and descriptions. To focus on career-track employment, we remove jobs that occurred during a user's first continuous education period after high school, such as internships. For example, an undergraduate degree immediately followed by a master's degree within the same calendar year is treated as one continuous education period, and internships during this period are removed from our analysis. After filtering, we analyze over 7.5 million work experiences with end dates after 2018, yielding rich longitudinal data on career trajectories.

\subsection{Analysis Window Selection}
We focus on the data in the period from June 2021 to June 2024, and conduct formal statistical analysis from June 2022 to June 2024. These time windows are chosen for several reasons. First, June 2021 marks the point when labor markets began stabilizing after the acute phase of the COVID-19 pandemic. According to the U.S. Bureau of Labor Statistics, the unemployment rate stabilized around 5.9\%, marking a transition toward post-pandemic recovery in labor markets~\cite{bls2021employment}. Second, according to the REALM dataset, LLM-related discussion volumes on Reddit and in news articles were minimal before mid-2021, providing insufficient variation for meaningful analysis. Third, we begin formal statistical analysis in June 2022 to avoid a structural break visible in the job posting index around this time (Figure~\ref{fig:disc_posting}). This turning point likely reflects the confluence of Federal Reserve monetary tightening~\cite{smialek2022fedrates}, and broader tech sector hiring corrections in June 2022~\cite{metz2022hiringfreezes, mickle2022techlayoffs}. Therefore, we start the analysis after June 2022 transition period to avoid conflating multiple macroeconomic shocks with GenAI-specific effects or potentially violating stationarity assumptions. Most importantly, our analysis window (June 2022 to June 2024) encompasses the launch of major LLM products that catalyzed widespread public awareness and adoption, most notably ChatGPT (November 2022)\footnote{https://openai.com/blog/chatgpt}, GPT-4 (March 2023)\footnote{https://openai.com/gpt-4-research}, and other prominent models, ensuring our study captures the period when GenAI discussion became more substantial. By analyzing this post-June 2022 period, we focus on a more stable macroeconomic environment where the relationship between GenAI discussion and employment dynamics can be identified more cleanly.

\subsection{Data Preprocessing Methods}

We apply two preprocessing steps to prepare the data for analysis. First, we classify job postings and work experience by occupation using O*NET occupation codes, maintaining consistency with how REALM processes LLM discussion data from Reddit and news articles. This allows us to examine GenAI's impact across occupational domains and assess whether online discussions serve as leading indicators of labor market changes. Second, we train a binary classifier to identify whether job postings or work experience in the LinkedIn dataset involve GenAI-related work. For both types of classification, manual annotation is required. The criteria are described below. These preprocessing steps ensure consistency in occupational categorization and GenAI relevance across all datasets, enabling systematic cross-dataset analysis.


\subsubsection{\textbf{Occupation and GenAI-Related Classification for Job Postings}}

\paragraph{\textbf{Occupation Classification:}}
Job postings include occupation labels in the raw data, requiring no additional preprocessing.

\paragraph{\textbf{GenAI-Related Classification:}}
We annotate a job posting as GenAI-related if it satisfies one of the following criteria:\\
\noindent\textbf{(1) Direct GenAI Roles:} The posting explicitly requires knowledge, skills (\emph{e.g.}, fine-tuning, prompt engineering, model evaluation), or experience with specific LLMs or related tools (\emph{e.g.}, LangChain~\cite{mavroudis2024langchain}). This category also includes positions focused on LLM-specific infrastructure, such as training or inference pipelines, model deployment at scale, memory optimization, or latency reduction.\\
\noindent\textbf{(2) Supporting GenAI Roles:} The posting involves work that supports LLM-based products or applications, even when LLMs are not explicitly mentioned. For instance, a data engineer position developing real-time pipelines for processing unstructured financial documents to feed an LLM-powered financial application would qualify based on this criterion.

Since GenAI-related job postings represent only a small portion of all postings, we used stratified sampling to create a balanced training set. We manually selected 450 postings likely to be GenAI-related from categories such as machine learning engineer and data analyst, then supplemented these with 1,550 postings randomly sampled from all occupations. Then we supplemented with 1,550 postings randomly sampled from all occupations. Two annotators independently labeled these datapoints using the aforementioned two criteria, resolving disagreements through three structured discussion sessions to ensure consistency. This process yielded 1,600 training samples and 400 validation samples. We fine-tuned a Qwen3-4B model~\cite{yang2025qwen3} on this annotated dataset, achieving 97.5\% accuracy and an F1 score of 87.5\% on the held-out validation set, demonstrating strong classification performance for this task.

\subsubsection{\textbf{Occupation and GenAI Classification for LinkedIn Profiles}}

The data preprocessing step for LinkedIn profiles focuses on standardizing each user’s work experience records. For every experience, we identify the associated occupation and determine whether that experience is related to GenAI.

\paragraph{\textbf{Occupation Classification:}}
Given the LinkedIn dataset's 4 million work experiences, directly applying a fine-tuned LLM classifier is computationally infeasible. We adopted a more efficient approach. Two annotators manually labeled 3,500 job postings across 13 O*NET occupational categories, conducting structured discussion sessions to ensure consistency. Rather than training a language model, we used an MPNet sentence transformer (all-mpnet-base-v2) to convert position titles and descriptions into embeddings, then trained a multinomial logistic regression model for multi-class classification. On a held-out test set of 1,400 samples, the model achieved 73.7\% accuracy and a macro F1 score of 74.4\%. This level of accuracy is sufficient for large-scale analysis, as aggregate patterns across 4 million work experiences are robust to individual classification errors.

\paragraph{\textbf{GenAI-Related Classification:}}
Unlike job postings, work experience descriptions in LinkedIn profiles tend to be concise and direct: users usually explicitly mention relevant technologies when applicable, which makes keyword matching particularly effective. Thereby we use a simple rule-based approach to identify GenAI-related work experiences: \\
(1) \emph{Keyword matching:} We flag roles whose descriptions contain terms such as ``LLM,'' ``fine-tuning,'' or ``prompt engineering'' (see Appendix~\ref{appx:genai_labeling} for the complete keyword list).  \\
(2) \emph{Company-level identification:} We label roles at companies focused exclusively on GenAI (\emph{e.g.}, OpenAI, Anthropic, Mistral AI) as GenAI-related (see Appendix~\ref{appx:genai_labeling} for the company list). 

A work experience is classified as GenAI-related if either criterion is met. This dual strategy efficiently captures GenAI-related roles at scale.

\section{Metric Computation}
We address our two research questions through a systematic analysis of labor market dynamics. 
To examine the relationship between online discussions and labor market changes, we develop detailed, timely, and granular metrics that capture various dimensions of occupational mobility: transitions into and out of different job domains, tenure duration within occupations, job search duration, and movement toward GenAI-related positions. 
We apply these metrics to observational data to Indeed job postings and LinkedIn user profiles, measuring monthly changes to capture labor market dynamics stratified by occupation.

Our approach complements existing national labor surveys.\footnote{https://www.bls.gov/bls/descriptions.html} While surveys conducted by the Census Bureau and the Bureau of Labor Statistics (\emph{e.g.}, Current Population Survey, American Community Survey, Job Openings and Labor Turnover Survey, Occupational Employment and Wage Statistics) provide authoritative statistics on employment, unemployment, wages, and job flows, they face important limitations for our purposes. These surveys are generally designed to capture aggregate patterns at the national or industry level, are conducted at annual or quarterly frequencies, and rely on representative samples of households or employers. While invaluable for population-level benchmarking, these characteristics make official statistics less suitable for tracking occupation-specific, high-frequency responses to emerging technological shocks such as GenAI adoption.

\paragraph{\textbf{Net Change Ratio (NCR)}}
Conceptually similar to the hires and separation rates used in the National Labor Survey\footnote{https://www.bls.gov/news.release/jolts.html}, this metric captures the job inflows and outflows within an occupation, reflecting whether the occupation is expanding or contracting. It is defined as the net difference between the number of people starting in this occupation and the number of people leaving this occupation, normalized by the total active workforce in that occupation:
\begin{equation}
\text{NCR}_{t}^{(occ)} = 
\frac{\text{Starts}_{t}^{(occ)} - \text{Ends}_{t}^{(occ)}}{\text{Active}_{t}^{(occ)}},
\end{equation}
where $\text{Active}_{t}^{(occ)}$ is the total active workforce in occupation $occ$ at time $t$. Positive values indicate occupational growth, while negative values suggest contraction.

\paragraph{\textbf{Normalized Tenure}}
To measure job stability while accounting for right-censoring in our data, we calculate the average tenure of jobs normalized by the maximum possible tenure \emph{i.e.,} the time elapsed between job start and data collection. We use monthly granularity for this metric:
\begin{equation}
\text{Tenure}_{t}^{(occ)} = 
\frac{1}{N_{t}^{(occ)}} \sum_{i=1}^{N_{t}^{(occ)}} 
\frac{\text{months employed}_i}{\text{max possible tenure}},
\end{equation}
where $N_{t}^{(occ)}$ is the number of job starts in occupation $occ$ during month $t$, and $\text{max possible gap} = \text{current month} - \text{start month}_i$.
This normalization corrects for right-censoring bias, since jobs starting recently have shorter potential tenure to observe. Hence, normalization enables consistent comparisons across different time periods. Higher values indicate that workers are remaining longer in their positions relative to expectations.


\paragraph{\textbf{Unemployment Duration}}
This metric measures how quickly workers enter specific occupations following job loss. For each job transition, we calculate the duration between job termination and the start of the next position, grouped by the occupation of the subsequent role and normalized by the maximum observable gap to address right-censoring. We use monthly granularity:
\begin{equation}
\text{UnempDur}_{t}^{(occ)} =
\frac{1}{M_t^{(occ)}} \sum_{j=1}^{M_t^{(occ)}}
\frac{\text{gap months}_j}{\text{max possible gap}_j}.
\end{equation}
where $M_{t}^{(occ)}$ is the number of job ends in occupation $occ$ during month $t$, and $\text{max possible gap} = \text{current month} - \text{end month}_j$.
This metric captures how quickly workers enter a given occupation after job loss. Higher values indicate longer job search durations for workers entering that occupation.

\paragraph{\textbf{GenAI Transition Ratio}}
To capture the rate at which workers move into AI-focused positions, we define a GenAI transition as a job change where the prior position is non-GenAI-related and the subsequent position is GenAI-related. The transition ratio is computed as:
\begin{equation}
\text{GenAI Trans}_{t} =
\frac{\text{Transitions into GenAI}_{t}}{\text{All new starts}_{t}}.
\end{equation}
This metric tracks the proportion of new job starts that represent movement into GenAI roles, providing insight into the growing penetration of AI across the workforce.

To address potential censoring effects and capture varying temporal dynamics, we additionally compute tenure and unemployment duration distributions using categorical time buckets (0–3 months, 4–12 months, 12+ months). This approach provides granular insights into short-term versus long-term patterns.
Together, these four metrics enable us to examine heterogeneous, occupation-specific labor market responses and link shifts in online LLM discourse to concrete employment dynamics at a monthly resolution.

In the following sections, we address our two research questions through a systematic analysis of labor market dynamics. First, we examine whether GenAI-related employment has grown over time and compare tenure and unemployment patterns between workers in GenAI roles and those in non-GenAI occupations to assess how AI adoption influences labor market outcomes (RQ1; §\ref{sec:observation}). Next, we investigate whether online discussion activity serves as a leading indicator of labor market change (RQ2) by analyzing four key employment metrics: job postings (§\ref{sec:jobposting}), net change ratio (§\ref{sec:NCR}), tenure (§\ref{sec:tenure}), and unemployment duration (§\ref{sec:unemploy}).


\section{RQ1: Understanding Impact}

In this section, we examine the influence of GenAI on the labor market by comparing employment patterns between workers in GenAI-related roles and those in non-GenAI roles. We begin with empirical observations of GenAI adoption and labor demand trends across occupations, then conduct formal statistical comparisons of employment dynamics between the two cohorts.

\subsection{Observations}
\label{sec:observation}

\paragraph{\textbf{GenAI Adoption Patterns}}
\begin{figure}[htbp]
    \centering
    \includegraphics[width=1 \linewidth]{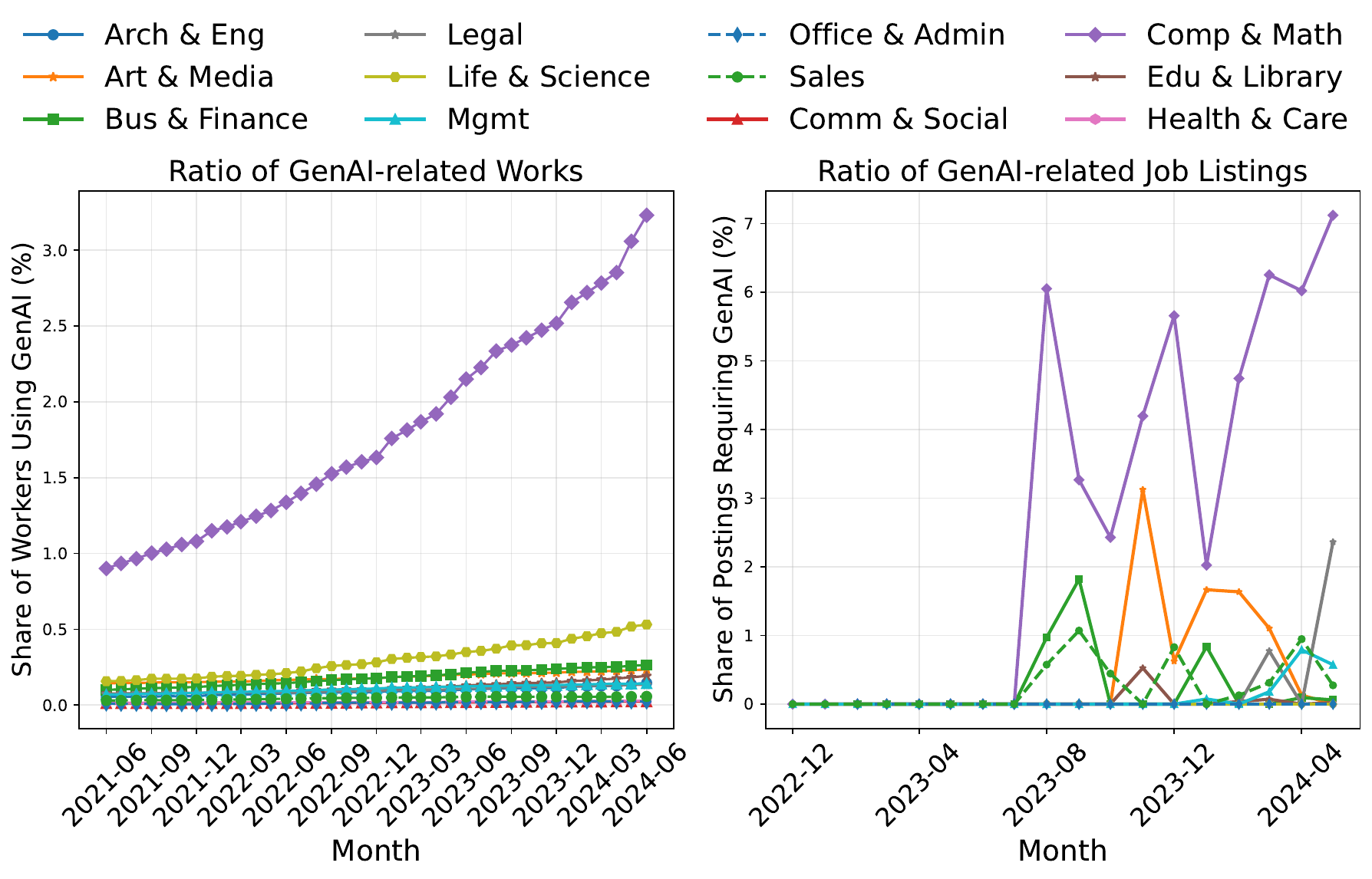}
    \caption{Trends in GenAI adoption by occupation: Monthly share of workers using GenAI skills within each occupation (left) and monthly share of job listings requiring GenAI skills within each occupation (right).}
    \label{fig:GenAI_exp_listing}
\end{figure}

Figure~\ref{fig:GenAI_exp_listing} reveals a clear increasing trend in GenAI-related employment within the Computer \& Math occupation. For workers using GenAI skills (left panel), the Computer \& Math sector shows steady, consistent growth from mid-2021 to mid-2024, indicating that GenAI skills have become increasingly prevalent among technical workers. The job listings data (right panel), while limited by data sparsity prior to December 2022 due to the scraping methodology, reveals a similar upward trend in the Computer \& Math occupation for positions requiring GenAI skills. This pattern demonstrates that GenAI integration has been highly concentrated within technical occupations, with Computer \& Math showing the most substantial and sustained within-occupation adoption over time.

\paragraph{\textbf{Labor Demand Trends}}
\begin{figure}[htbp]
    \centering
    \includegraphics[width=1 \linewidth]{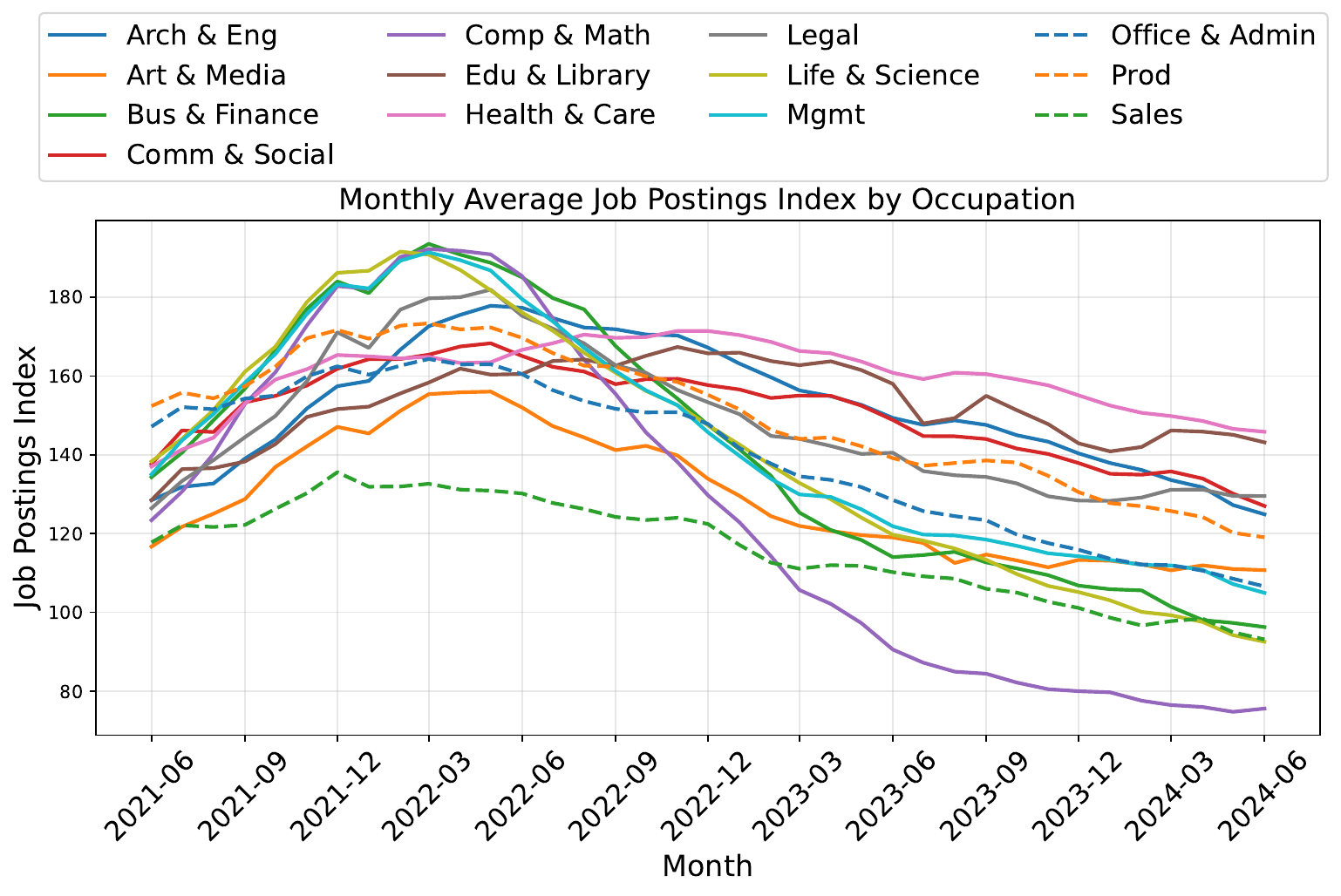}
    \caption{Trends in job posting index by occupation.}
    \label{fig:disc_posting}
\end{figure}

Despite increasing GenAI adoption, overall labor demand has decreased. As shown in Figure~\ref{fig:disc_posting}, job postings across all occupations exhibit a downward trend after mid-2022. The decline is most pronounced for Computer occupations, which experience the sharpest drop, whereas occupations such as Healthcare exhibit relatively steadier patterns with smaller decreases. This divergence between rising GenAI skill requirements and declining overall job postings suggests a transformation rather than an expansion of labor demand in technical fields.

\subsection{Comparing GenAI and Non-GenAI Employment Dynamics}
\label{sec:GenAI_compare}

To systematically examine GenAI's impact on labor market outcomes, we directly compare employment patterns between GenAI-related and non-GenAI jobs. We focus on two key metrics constructed from LinkedIn profile data: job tenure and unemployment duration. The \textbf{GenAI} cohort consists of workers who have ever held a GenAI-related job; for these workers, we retain only jobs \emph{after} their first GenAI position to isolate the effect of GenAI experience. The \textbf{non-GenAI} cohort includes all other workers.

\subsubsection{Statistical Methods}
We employ two complementary procedures to quantify differences between the cohorts. First, for each outcome $b$, we estimate an Ordinary Least Squares (OLS) regression with a group indicator and a group--time interaction:
\begin{equation}
\label{eq:trenda}
Y_{b,g,t}
\;=\;
\alpha_b
\;+\;
\beta_b\,\mathbf{1}[g=\text{GenAI}]
\;+\;
\gamma_b\,t
\;+\;
\delta_b\,\mathbf{1}[g=\text{GenAI}]\,t
\;+\;
\varepsilon_{b,g,t},
\end{equation}
where $Y_{b,g,t}$ is the outcome measure for metric $b$, group $g$ (GenAI or non-GenAI), at time $t$. The intercept $\alpha_b$ represents the baseline level for the non-GenAI group, while $\gamma_b$ captures the general time trend. We summarize results with (i) the \emph{baseline level gap} $\widehat{\beta}$, defined as the GenAI minus non-GenAI difference at $t=0$, and (ii) the \emph{monthly slope difference} $\widehat{\delta}$, the difference in linear trends per month between the two groups. A significant $\widehat{\beta}$ indicates a level gap at the start of the observation window; a significant $\widehat{\delta}$ indicates divergence ($\widehat{\delta}>0$) or convergence ($\widehat{\delta}<0$) over time.

Second, to test for differences in the entire time paths beyond linear trends, we use a dynamic time-warping (DTW) permutation test with $B=2000$ random relabelings. The reported $p$-value represents the proportion of permuted DTW distances that equal or exceed the observed distance, providing a non-parametric assessment of trajectory similarity.

\subsubsection{Results}

\begin{figure}[htbp]
    \centering
    \includegraphics[width=\linewidth]{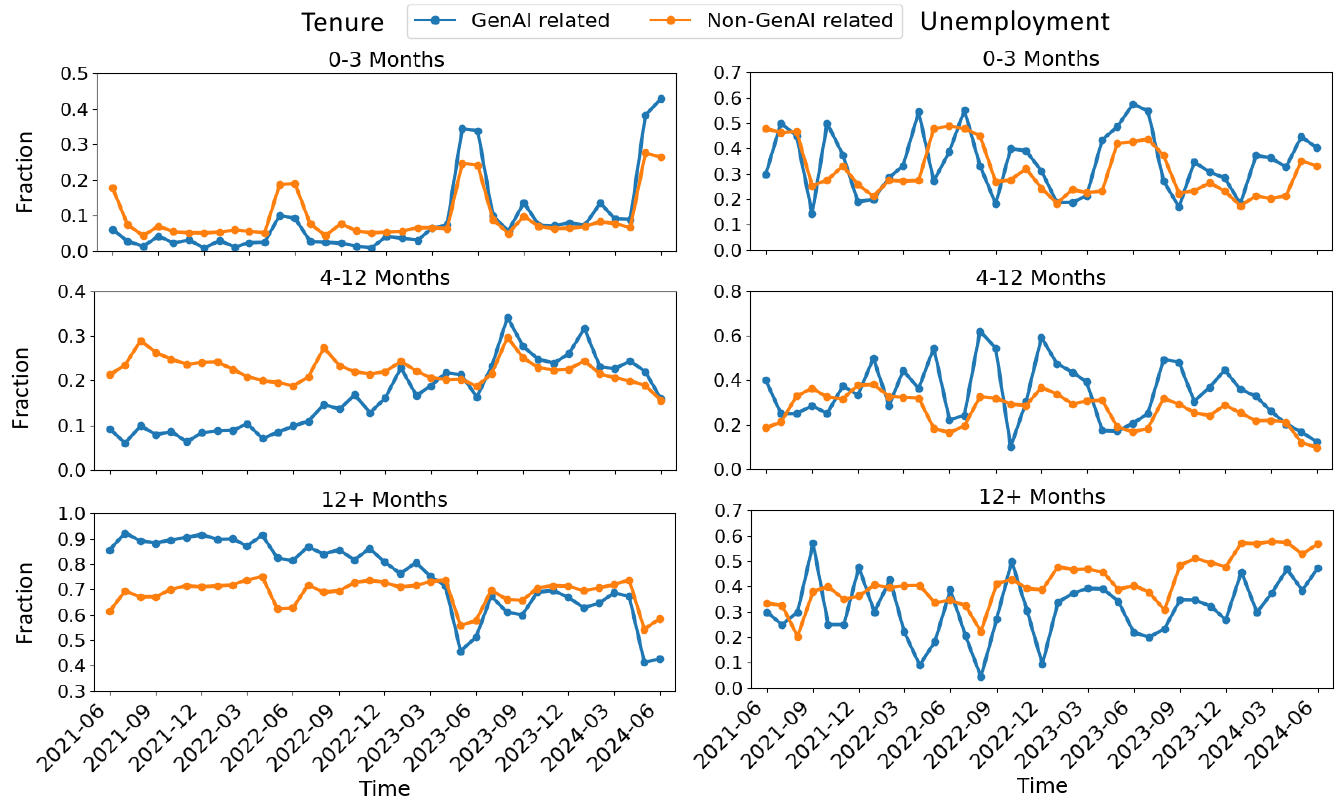}
    \caption{Comparison of tenure (left) and unemployment durations (right) of different lengths between GenAI versus Non-GenAI groups. Orange/blue denote GenAI/Non-GenAI groups respectively.}
    \label{fig:GenAI_compare}
\end{figure}

\begin{table}[t]
\centering
\caption{Regression estimates and DTW permutation statistics comparing GenAI and non-GenAI workers across tenure and unemployment bins.}
\small
\begin{tabular}{l|cc|c}
\hline
& \multicolumn{2}{c}{Regression} & \multicolumn{1}{|c}{DTW test} \\
\cline{2-3}\cline{4-4}
Outcome bin & Baseline gap $\widehat{\beta}$ & Slope diff.\ $\widehat{\delta}$ & DTW stat. \\
\hline
\multicolumn{4}{l}{\textit{Panel A: Tenure fractions}} \\
0--3 months   & -0.058              &  0.0057               & 1.664          \\
4--12 months  & -0.090$^{***}$      &  0.0063$^{***}$       & 2.978          \\
12+ months    &  0.151$^{***}$      & -0.0119$^{***}$       & 3.189$^{***}$  \\
\midrule
\multicolumn{4}{l}{\textit{Panel B: Unemployment fractions}} \\
0--3 months   & -0.016              &  0.0052               & 2.879          \\
4--12 months  &  0.097              & -0.0014               & 2.961$^{**}$   \\
12+ months    & -0.081$^{*}$        & -0.0038               & 3.422$^{**}$   \\
\hline
\end{tabular}

\vspace{0.3ex}
\parbox{0.92\linewidth}{\footnotesize
Notes: 
${}^{*}p<0.10$, ${}^{**}p<0.05$, ${}^{***}p<0.01$.
}
\label{tab:regression_dtw_combined}
\end{table}

\paragraph{\textbf{Tenure}}
The key observation is that GenAI workers shift toward medium tenure and away from long tenure over time compared to non-GenAI workers. Figure~\ref{fig:GenAI_compare} (left panel) plots monthly shares across three tenure bins (0--3 months, 4--12 months, 12+ months).
Specifically, in the 4--12 month bin, GenAI workers start at a lower share than non-GenAI workers but grow faster:
$\widehat{\beta}_{\text{4--12m}}=-0.0898$ ($p<0.001$) and
$\widehat{\delta}_{\text{4--12m}}=0.0063$ per month ($p=0.002$).
In the 12+ month bin, GenAI workers start at a higher share but decline more quickly:
$\widehat{\beta}_{\text{12+m}}=0.1507$ ($p<0.001$) and
$\widehat{\delta}_{\text{12+m}}=-0.0119$ per month ($p<0.001$).
The DTW test confirms a distinct trajectory for the 12+ month bin ($p=0.001$), consistent with the strong negative slope for GenAI workers.

\paragraph{\textbf{Unemployment Duration}}
The key observation is that GenAI workers experience faster re-employment after job separation, with significantly lower exposure to long-term unemployment and different medium- and long-duration adjustment paths over time, compared to non-GenAI workers. Figure~\ref{fig:GenAI_compare} (right panel) shows unemployment duration shares across three bins (0--3 months, 4--12 months, 12+ months). 
Specifically, for long-term unemployment (12+ months), GenAI workers begin at a lower level:
$\widehat{\beta}_{\text{12+m}}=-0.0812$ ($p<0.10$), while the slope difference is small and not significant.
DTW permutation tests detect trajectory differences for both 4--12 months ($p<0.05$) and 12+ months ($p<0.05$).

\paragraph{\textbf{Summary}}
Taken together, the evidence reveals a distinct employment reallocation pattern for GenAI workers following their first GenAI job: the medium-tenure (4--12 months) share increases relative to non-GenAI workers, while the long-tenure (12+ months) share decreases. Additionally, GenAI workers exhibit lower levels of long-duration unemployment and different trajectories across unemployment duration categories, suggesting more dynamic labor market mobility and faster job-to-job transitions.

\section{RQ2: Discussion as Leading Indicator}

In this section, we investigate whether fluctuations in online discussion intensity about LLMs can anticipate changes in labor market outcomes. We begin by examining descriptive patterns that motivate our focus on online discussion, then apply formal statistical methods, including Granger causality tests \cite{shojaie2022granger, seth2007granger, friston2014granger} and out-of-sample prediction analyses \cite{tsamardinos2018bootstrapping, borup2022anatomy}, to assess the extent to which discussion signals have predictive power for subsequent shifts in employment metrics.

\subsection{Observations}
\begin{figure}[htbp]
    \centering
    \includegraphics[width=1 \linewidth]{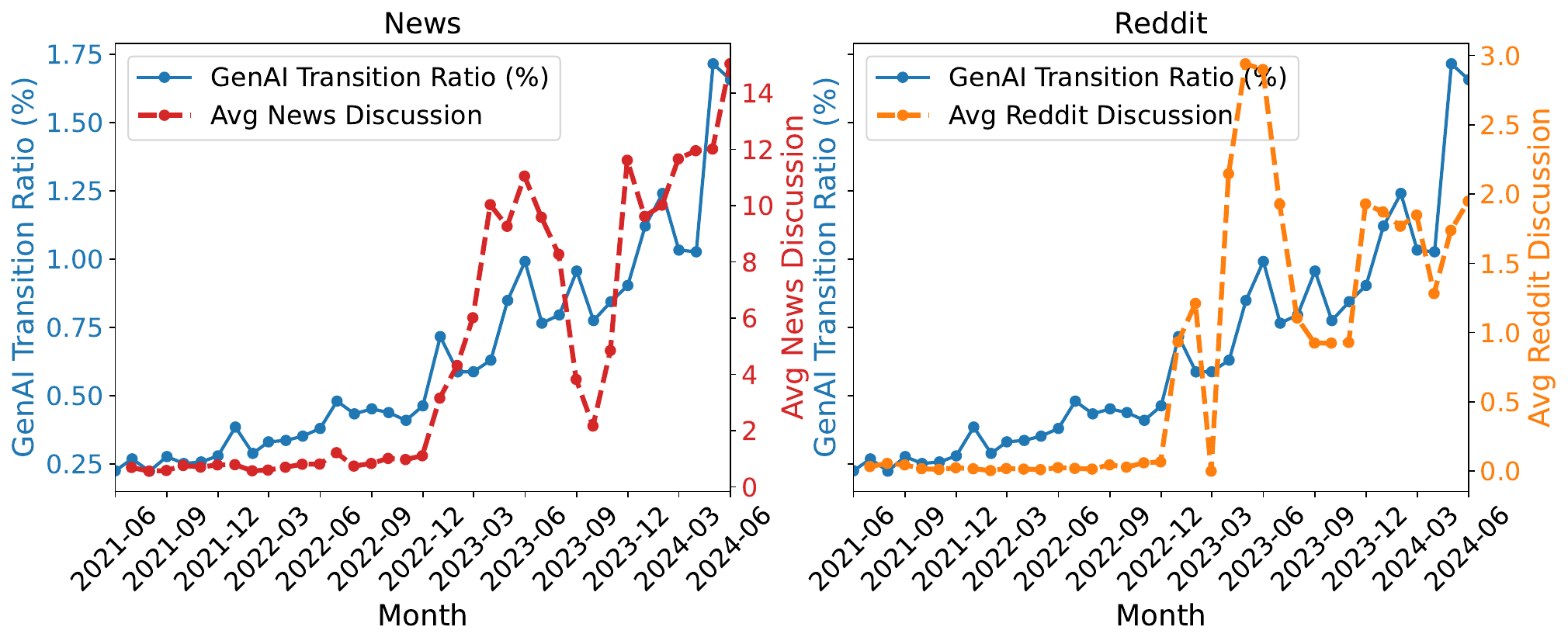}
    \caption{GenAI transition patterns and online discussion intensity over time. The GenAI transition ratio represents the monthly fraction of workers starting GenAI-related jobs from non-GenAI positions among all job starts. Discussion metrics show the intensity of LLM-related discourse in News (left) and Reddit (right) platforms.}
    \label{fig:transition_metric}
\end{figure}

Figure~\ref{fig:transition_metric} reveals suggestive temporal relationships between online discussion intensity and GenAI labor market transitions. Both News and Reddit exhibit dramatic spikes in LLM-related discussion during 2022-2023, alongside general upward trends over time. Notably, these discussion surges precede and coincide with substantial increases in GenAI transition ratios: the proportion of workers moving from non-GenAI to GenAI positions. The timing suggests that online discourse may serve as a leading indicator for career transitions into GenAI roles, motivating our formal statistical investigation below.

\subsection{Statistical Methods Used for Leading Indicator Analysis}
We employ two complementary approaches to assess whether online discussion activity could serve as a leading indicator for labor market changes. 


\paragraph{\textbf{Granger Causality}}
We test whether past discussion volumes contain information useful for predicting labor market metrics beyond what is contained in the metrics' own historical values. This involves comparing restricted autoregressive (AR) models using only lagged outcome variables against unrestricted models that include both lagged outcomes and lagged discussion variables. The null hypothesis that discussion does not Granger-cause labor market changes is tested using $F$-statistics. A significant p-value (\emph{i.e.,} smaller than 0.05) rejects the null hypothesis and indicates that discussion activity provides predictive information about future labor market conditions.

\paragraph{\textbf{Out-of-Sample Prediction}}
To assess practical forecasting value, we compare the predictive accuracy of baseline AR models against augmented models that include discussion variables. Using a rolling window approach with 10-month training periods, we generate 1- and 3-month ahead forecasts and evaluate performance using mean squared forecast errors. We compute out-of-sample $R^2$ values to measure the percentage improvement in forecast accuracy from including discussion data. Statistical significance is assessed using Clark-West tests for nested model comparisons. A positive and higher $R^2$ value indicates that discussion variables meaningfully improve the prediction of labor market outcomes.

In the following subsections, we apply these methods to analyze four labor market indicators: total job postings, net change ratios (NCR), tenure duration, and unemployment duration.

\subsection{Total Job Postings}
\label{sec:jobposting}
\begin{figure}[htbp]
    \centering
    \includegraphics[width=1 \linewidth]{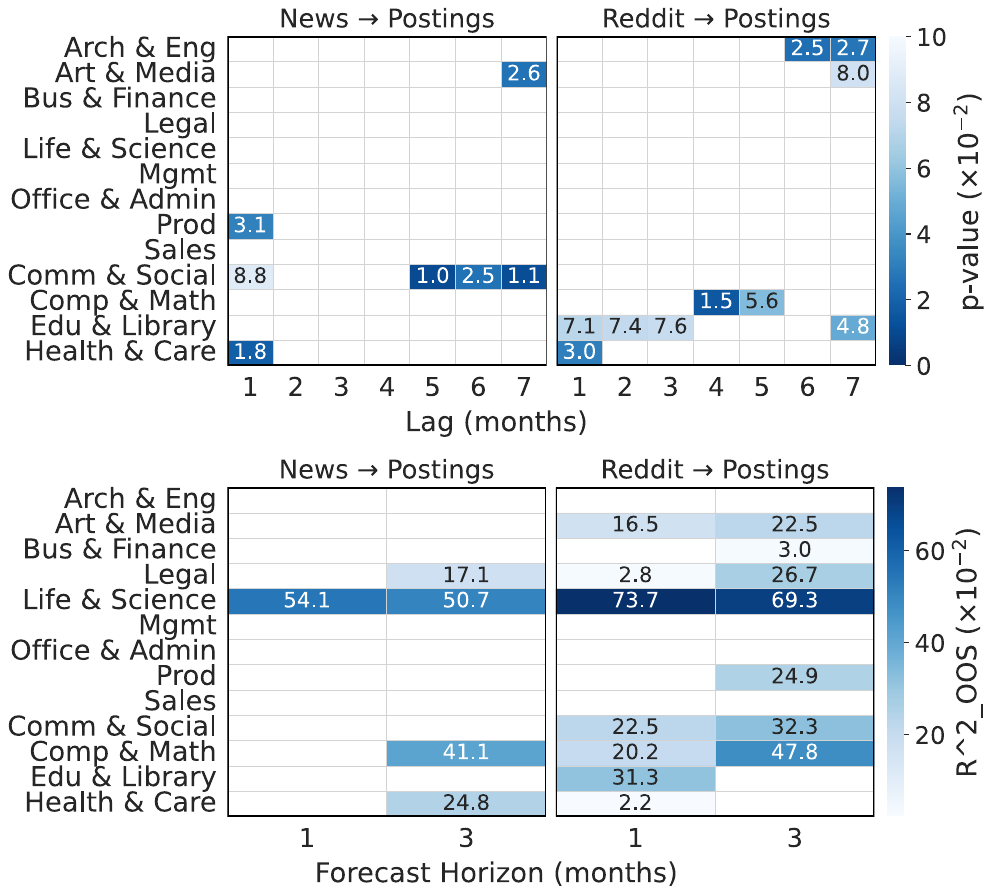}
    \caption{Granger causality (top) and out-of-sample prediction (bottom) of the job posting index from online discussion. 
    The top panels report Granger causality test statistics (p-values) for the predictive relationship between news (left) and Reddit (right) discussion intensity and job postings at different lags. Only significant values ($p < 0.01$) are shown, where significance indicates that discussion Granger-causes changes in the job posting index. 
    The bottom panels present out-of-sample prediction accuracy at 1- and 3-month forecast horizons. We display $R^2_{\text{OOS}}$ values only when their significance level is below 0.05. Positive values highlight that the online discussion has predictive power on the job posting index across occupational categories.}
    \label{fig:job_posting_metric}
\end{figure}


\textbf{Granger-Causality.}
The results reveal heterogeneous response speeds across occupations (Figure~\ref{fig:job_posting_metric}, top panels). Healthcare exhibits rapid responses: discussion changes predict job posting changes within one month on both News and Reddit platforms. In contrast, creative and education occupations adjust more slowly, with predictive effects emerging around 7 months on Reddit (Architecture, Arts, Education) and at similar horizons on News (Community, Arts). Computer occupations also demonstrate delayed responses, with significant predictive power at a 5-month lag on Reddit.
\\
\textbf{Out-of-Sample Prediction.} Reddit improves short-horizon (1-month) forecasts broadly across occupations, including Arts, Community, Computer, Education, Healthcare, Legal, Life Sciences, and Production (Figure~\ref{fig:job_posting_metric}, bottom panels). News contributes primarily at the 3-month horizon, particularly for Computer, Healthcare, Legal, and Life Sciences. Life Sciences and Legal benefit from both sources, while Healthcare gains predictive accuracy at different horizons (Reddit at 1-month, News at 3-month).

Overall, the job posting index responds rapidly in Healthcare but slowly in creative occupations and Education. Reddit carries a broader short-horizon signal, while News adds value at longer forecasting horizons.

\subsection{Net Change Ratio (NCR)}
\label{sec:NCR}

\textbf{Granger-Causality.} 
NCR exhibits the clearest evidence of causal links. Reddit shows consistent short-lag effects across Architecture, Arts, Business, Computer, Education, and Management (mainly lags 1–3, occasionally extending to 5–6). News produces fewer signals but in broadly the same occupations (Architecture, Arts, Business, Management), again concentrated at 1–3 months. These results suggest that discussion shocks typically precede shifts in net hiring flows within a one-to-three-month window.
\\
\textbf{Out-of-Sample Prediction.}
Out-of-sample forecasts confirm that discussion intensity contains predictive information for hiring balance. Business shows consistent improvements from both sources at both the 1- and 3-month horizons. Arts exhibits gains at different horizons depending on the source, while Management benefits specifically from News at both 1 and 3 months.

Discussion activity, particularly on Reddit, serves as a leading indicator of short-term hiring dynamics. In particular, Business and Management occupations exhibit strong and stable predictability.
\begin{figure}[htbp]
    \centering
    \includegraphics[width=1 \linewidth]{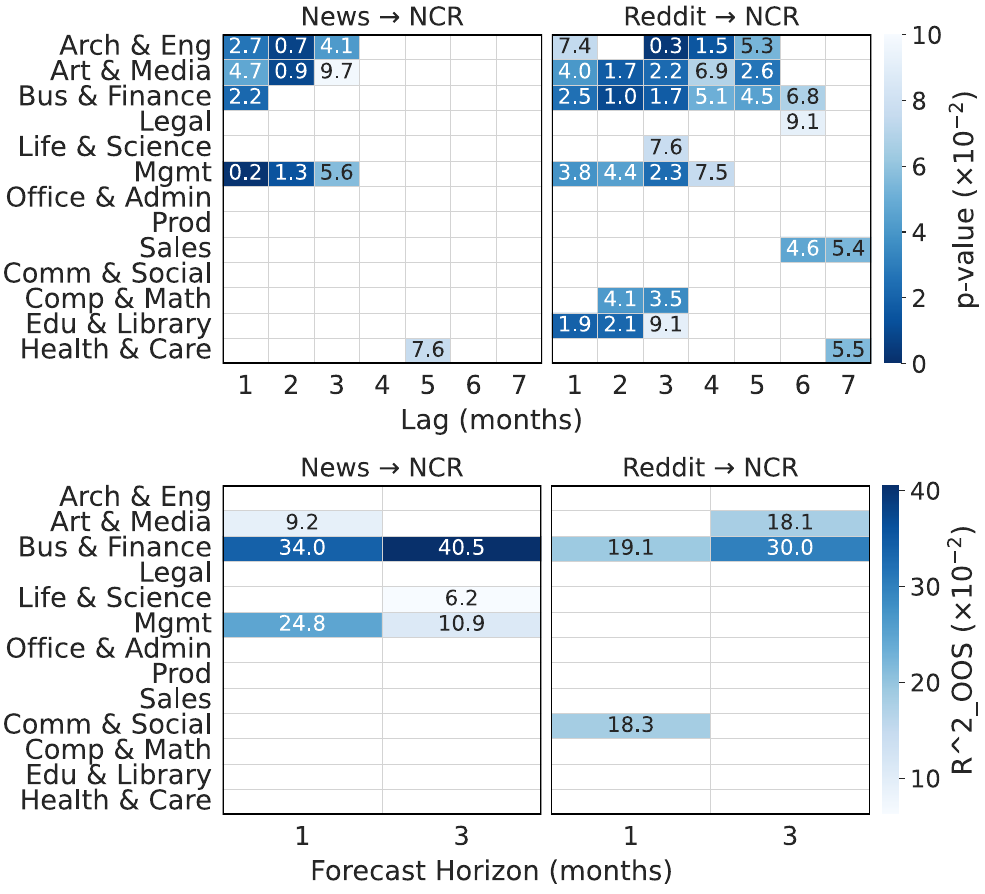}
    \caption{Granger causality (top) and out-of-sample prediction (bottom) of the net change ratio from online discussion (left: news, right: Reddit). }
    \label{fig:NCR_metric}
\end{figure}
\subsection{Tenure}
\label{sec:tenure}

\begin{figure}[htbp]
    \centering
    \includegraphics[width=1 \linewidth]{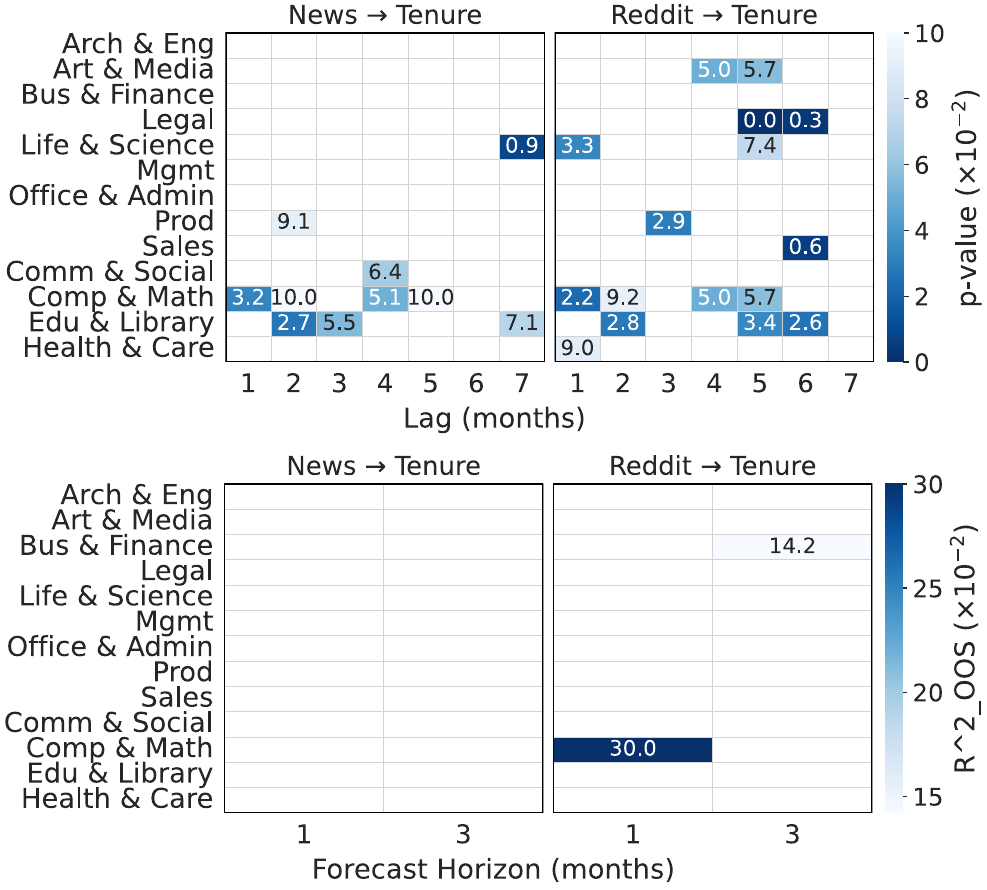}
    \caption{Granger causality (top) and out-of-sample prediction (bottom) results of the normalized tenure from online discussion (left: news, right: Reddit). }
    \label{fig:tenure_metric}
\end{figure}

\textbf{Granger-Causality.} 
Predictive signals are most consistent in Computer and Education occupations (Figure~\ref{fig:tenure_metric}, top panels). In Computer, both Reddit and News show significant effects at multiple short-to-medium lags (1, 2, 4, 5 months), indicating a robust temporal link between discussion shifts and subsequent changes in job retention patterns. Education also exhibits overlap across sources, with lag 2 significant in both. While Reddit additionally shows effects at lags 5–6 and News at lags 3 and 7, the shared lag 2 effect highlights a short-term relationship where discussion intensity precedes changes in the retention of new hires.
\\
\textbf{Out-of-Sample Prediction.}
Prediction results confirm that the Reddit discussion carries a predictive signal in Computer.

These results suggest that discussion intensity anticipates retention dynamics, whether new hires leave quickly or remain longer. In Computer, this likely reflects high churn in tech jobs, with workers responding rapidly to shifting demand or skill signals. In Education, the consistent lag~2 may capture shorter contracts or semester-driven employment cycles.

\subsection{Unemployment Duration}
\label{sec:unemploy}

\textbf{Granger-Causality.}
For unemployment duration, the strongest and most consistent effects appear at short horizons for Arts and Education, which show significant lags around 1–2 months in both sources, indicating that discussion shifts are quickly followed by changes in how long individuals remain unemployed before re-entering work in those occupations. Beyond these, Reddit picks up additional links in Architecture (lag 4), Computer (lag 6), Life (lag 3), and Sales (a lasting effect, lags 1–6). News emphasizes different areas, with signals for Business (lag 1), Healthcare (lags 2–3), Legal (lag 5), Life (lag 5), and Sales (lag 7). These patterns suggest that unemployment responses are occupation-specific: some fields adjust quickly, while others show longer reaction times between discussion and re-entry.
\\
\textbf{Out-of-Sample Prediction.}
Out-of-sample forecasts confirm predictive signal for Arts and Education. Reddit also improves longer-horizon predictions for Business, Healthcare, and Life (3 months). Education benefits from both sources, but at different horizons.

Unemployment responses are heterogeneous across occupations. Fast adjustments (Arts, Education) suggest that discussion intensity is informative about rapid re-entry, while slower responses point to longer adjustment cycles where discussion precedes re-employment only after several months.
\begin{figure}[htbp]
    \centering
    \includegraphics[width=1 \linewidth]{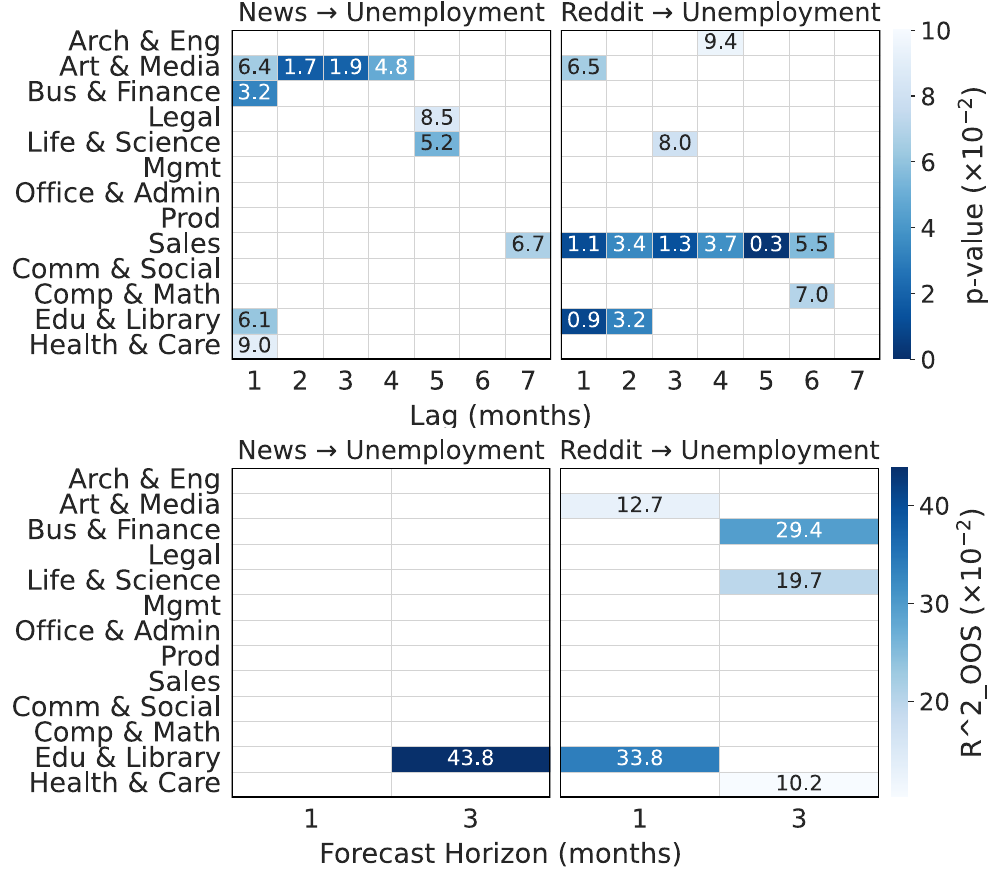}
    \caption{Granger causality (top) and out-of-sample prediction (bottom) of the unemployment duration from online discussion (left: news, right: Reddit). }
    \label{fig:unemployment_metric}
\end{figure}

\subsection{Summary}
Based on our analysis, three patterns stand out:

\paragraph{\textbf{Occupation-Specific Coverage}}
First, some occupations show significant results across metrics: Arts, Education, Computer, and Life. These occupations show significant results across all four metrics. These occupations demonstrate that online discourse captures a spectrum of labor market dynamics from hiring to unemployment duration. These are also knowledge-intensive occupations where workers actively engage in online professional discussions. Industry trends, technological changes, and policy shifts get debated online before affecting employment.

\paragraph{\textbf{Heterogeneous Response Timing}}
The speed at which labor markets respond to discussion changes varies systematically across both occupations and employment metrics. Most occupations exhibit short-term responses (1–3 months) for hiring dynamics (NCR) and unemployment duration, but longer delays (5–7 months) for job postings, particularly in creative occupations. However, substantial heterogeneity exists in occupation-specific response patterns. For job postings, Education and Healthcare respond within 1 month, while Architecture, Arts, and Computer show 5–7 month delays. NCR reveals a different pattern: Architecture, Arts, Business, Management, Education, and Computer all respond within 1–3 months, while Healthcare and Sales adjust more slowly. Tenure dynamics are fastest in Computer, Education, and Healthcare (1–2 months), but slower in Sales and Legal occupations (5+ months). Unemployment duration shows rapid responses in Arts, Business, Education, and Healthcare (1–2 months), with longer adjustments in Legal and Computer (5–6 months). 

Only two occupations demonstrate consistently fast responses across multiple metrics: Education (fast across all four metrics) and Healthcare (fast across three of four metrics). Most other occupations exhibit mixed response patterns. For instance, Arts responds quickly to unemployment changes (1–2 months) but slowly to job posting shifts (7 months), while Computer shows fast tenure responses (1–2 months) but slow job posting and unemployment adjustments (5–6 months). This heterogeneity suggests that different aspects of labor market adjustment follow distinct mechanisms that vary by occupational characteristics.

\paragraph{\textbf{Cross-Source Agreement}}
Clear cross-source agreement emerges for seven occupations: Education, Healthcare, Architecture, Arts, Business, Management, and Computer, where both Reddit and News show significant results at similar lags. This consistency suggests that these sectors generate labor market signals visible across different types of online discourse. However, the sources differ systematically in their coverage breadth and temporal focus. Reddit demonstrates broader coverage, showing significant effects across more occupations, particularly in out-of-sample prediction tests. In contrast, News covers fewer occupations but often provides value at longer forecasting horizons (3 months vs. 1 month). 

An example of this differential coverage appears in the Computer occupation: while Reddit shows significant predictive power across all four labor market metrics, News demonstrates significance for only one metric. This pattern suggests that Reddit more effectively captures the full spectrum of tech sector dynamics through active professional communities, while traditional news sources may miss some early signals.

This study demonstrates that online discussions about LLMs serve as reliable leading indicators of labor market changes, predicting employment dynamics 1-7 months in advance across multiple occupations, with the strongest predictive power for knowledge-intensive fields like Computer, Education, and Arts where workers actively engage in professional discourse online, revealing that digital conversations capture early signals of technological disruption before they materialize in formal employment statistics.


%

\section{Conclusion}
This study demonstrates that online discussions about LLMs serve as reliable leading indicators of labor market changes, predicting employment dynamics 1-7 months in advance across multiple occupations, with the strongest predictive power for knowledge-intensive fields like Computer, Education, and Art \& Media, and Life \& Social Science, where workers actively engage in professional discourse online, revealing that digital conversations capture early signals of technological disruption before they materialize in formal employment statistics. Additionally, we find that workers entering GenAI-related roles follow distinct career trajectories compared to their peers in non-GenAI roles. 

Beyond the empirical findings, our methodology highlights the potential of monitoring online discussions as a complement to conventional labor statistics, providing more timely insights into the interaction between technological innovation and labor market dynamics. For workers, this framework can help identify when to consider career plans. For organizations and policymakers, it offers a practical tool for anticipating shifts in the labor market and responding to disruptions.

\section{Acknowledgment}
This research was supported by funding from Open Philanthropy and benefited from data services provided by The Bright Initiative powered by Bright Data. We gratefully acknowledge their support in providing data used in our analysis.

\bibliographystyle{ACM-Reference-Format}
\bibliography{refs}

\appendix


\section{GenAI Identification}
\label{appx:genai_labeling}
To identify GenAI-related work experiences in the LinkedIn User Profile Dataset, we combine a company-based filter and a keyword-based filter.  
We designate positions as GenAI-related if they are associated with companies whose primary business centers on generative AI products or infrastructure. The company list is based on the 2025 Forbes \emph{Artificial Intelligence 50} \footnote{\url{https://www.forbes.com/lists/ai50/}} and includes well-known GenAI firms such as Anthropic, Cohere, Hugging Face, Mistral AI, OpenAI, and Scale AI, as well as specialized companies like Abridge, AnySphere, Baseten, Captions, Clay, Coactive AI, Crusoe, Decagon, DeepL, ElevenLabs, Figure AI, Fireworks AI, Glean, Harvey, Hebbia, Lambda, LangChain, Luminance, Mercor, Midjourney, Notion, OpenEvidence, Perplexity AI, Photoroom, Pika, Runway, Sakana AI, SambaNova, Skild AI, Snorkel AI, StackBlitz, Suno, Synthesia, Thinking Machine Labs, Together AI, Vannevar Labs, Vast Data, Windsurf, World Labs, Writer, and xAI.  

In addition, we label work experiences as GenAI-related if the job description contains terminology directly tied to generative AI models, methods, or applications. The keyword set covers:  
\begin{itemize}
    \item General AI terms (``generative AI,'' ``GenAI,'' ``foundation models'').  
    \item Model-specific names (``large language model,'' ``ChatGPT,'' ``GPT-3/4/5,'' ``LLaMA,'' ``Mistral,'' ``Claude,'' ``Gemini,'' ``DeepSeek'').  
    \item Company or product names (``OpenAI,'' ``Anthropic,'' ``LangChain,'' ``Cohere,'' ``Stability AI,'' ``Stable Diffusion,'' ``Midjourney,'' ``DALL-E'').  
    \item Technical roles and methods (``prompt engineering,'' ``prompt engineer,'' ``RLHF''/``reinforcement learning from human feedback,'' ``retrieval-augmented generation''/``RAG'').  
\end{itemize}
An experience is classified as GenAI-related if it either (i) belongs to one of the listed companies, or (ii) contains one or more of the keywords above.

\section{Analysis Method}

\subsection{Regression}
We estimate Eq.~\eqref{eq:trend} and summarize the results with (i) the \emph{baseline level gap} $\widehat{\beta}$, defined as the GenAI minus non-GenAI difference at $t=0$, and (ii) the \emph{monthly slope difference} $\widehat{\delta}$, the GenAI–non-GenAI difference in the linear trend per month. A significant $\widehat{\beta}$ indicates a level gap at the start of the window; a significant $\widehat{\delta}$ indicates divergence ($\widehat{\delta}>0$) or convergence ($\widehat{\delta}<0$) over time. 

For each outcome $b$, we estimate an OLS trend with a group indicator and a group--time interaction:
\begin{equation}
\label{eq:trend}
Y_{b,g,t}
\;=\;
\alpha_b
\;+\;
\beta_b\,\mathbf{1}[g=\text{GenAI}]
\;+\;
\gamma_b\,t
\;+\;
\delta_b\,\mathbf{1}[g=\text{GenAI}]\,t
\;+\;
\varepsilon_{b,g,t}.
\end{equation}
Here $\beta_b$ is the baseline gap at $t=0$ and $\delta_b$ is the monthly slope difference (GenAI minus non-GenAI).
We test $H_0\!:\beta_b=0$ (no baseline gap) and $H_0\!:\delta_b=0$ (no slope difference).

\subsection{Dynamic-time-warping (DTW)}
Second, we test for differences in the entire time paths using a dynamic time-warping (DTW) permutation test\cite{KIANIMAJD201711005} with $B=2000$ random relabelings; the reported $p$-value is the share of permuted DTW distances that are at least as large as the observed distance.
Let $\mathbf{y}^{\text{AI}}=(y^{\text{AI}}_1,\ldots,y^{\text{AI}}_T)$ and
$\mathbf{y}^{\text{Non}}=(y^{\text{Non}}_1,\ldots,y^{\text{Non}}_T)$ denote the two cohort series for a given outcome (optionally $z$-scored).
The DTW distance is computed by the standard recursion
\begin{align}
D(i,j)
&=
d\!\left(y^{\text{AI}}_i,\,y^{\text{Non}}_j\right)
+
\min\!\big\{D(i\!-\!1,j),\,D(i,j\!-\!1),\,D(i\!-\!1,j\!-\!1)\big\},
\\
D(1,1)&=d\!\left(y^{\text{AI}}_1,\,y^{\text{Non}}_1\right),
\qquad
D(i,0)=D(0,j)=+\infty,
\end{align}
with $d(u,v)=(u-v)^2$.
Let $d_{\text{obs}}=D(T,T)$.
We obtain a permutation $p$-value by randomly reassigning month-level observations to two groups (while keeping lengths $T$ and marginal distributions), recomputing the DTW distance $d^{(b)}$ for $b=1,\ldots,B$, and evaluating
\begin{equation}
\label{eq:perm}
p
\;=\;
\frac{1+\sum_{b=1}^B \mathbf{1}\!\left\{d^{(b)}\ge d_{\text{obs}}\right\}}
{1+B},
\qquad
B=2000.
\end{equation}





\subsection{Granger Causality}
\label{appx:causality}
Granger causality tests whether past values of one time series (X) contain information useful for predicting another series (Y) beyond what is contained in Y's own past values. The test is based on comparing restricted and unrestricted Vector Autoregressive (VAR) models.

For bivariate Granger causality testing, we estimate two models:

Restricted Model (AR):
\begin{equation}
\label{eq:AR}
Y_t = \alpha_0 + \sum_{i=1}^{p} \alpha_i Y_{t-i} + \varepsilon_t
\end{equation}

Unrestricted Model (ARDL):
\begin{equation}
\label{eq:ARDL}
Y_t = \beta_0 + \sum_{i=1}^{p} \beta_i Y_{t-i} + \sum_{i=1}^{p} \gamma_i X_{t-i} + u_t
\end{equation}
where $p$ is the lag length, $\varepsilon_t$ and $u_t$ are error terms.
The null hypothesis is $H_0: \gamma_1 = \gamma_2 = ... = \gamma_p = 0$ (X does not Granger-cause Y). We then perofrm the joint significance test: $$F = \frac{(RSS_R - RSS_U)/p}{RSS_U/(T-2p-1)}$$
where $RSS_R$ and $RSS_U$ are residual sum of squares from restricted and unrestricted models.

Under $H_0$, F follows an F-distribution with $(p, T-2p-1)$ degrees of freedom.
If the p-value is significant, that is evidence that discussion volumes Granger-cause the labor metric

\subsection{Out of Sample Prediction Analysis}
\label{appx:OOSprediction}

To evaluate the predictive value of online discussion measures, we implement out-of-sample forecasting. We compare a baseline autoregressive (AR) model with an augmented autoregressive distributed lag (ARDL) specification that incorporates discussion variables.

The baseline AR model is defined as
\begin{equation}
Y_{t+h} = \phi_0 + \sum_{i=1}^{p} \phi_i Y_{t-i+1} + \varepsilon_{t+h},
\end{equation}
while the augmented ARDL model takes the form
\begin{equation}
Y_{t+h} = \theta_0 + \sum_{i=1}^{p} \theta_i Y_{t-i+1} + \sum_{i=1}^{p} \delta_i X_{t-i+1} + u_{t+h},
\end{equation}
where $h \in \{1,3\}$ denotes the forecast horizon, $p$ is the autoregressive lag length selected using the Bayesian Information Criterion (BIC), $Y_t$ is the outcome series, and $X_t$ denotes the discussion-based predictors.  

Forecasts are generated using a rolling window procedure. At each step, we estimate the models with a training sample of 10 consecutive monthly observations and then generate $h$-step-ahead forecasts. The estimation window is then rolled forward by one month, and the procedure is repeated until the end of the sample is reached.  

Forecast accuracy is evaluated using the mean squared forecast error (MSFE), defined as
\[
MSFE_m = \frac{1}{P} \sum_{t=R}^{T-h} (Y_{t+h} - \hat{Y}_{t+h,m})^2,
\]
where $m \in \{\text{AR}, \text{ARDL}\}$ indexes the model, $P$ is the total number of forecasts, and $R$ is the starting point of the rolling window.  

To assess the relative predictive performance, we compute the out-of-sample $R^2$ statistic:
\[
R^2_{\text{OOS}} = 1 - \frac{MSFE_{\text{ARDL}}}{MSFE_{\text{AR}}}.
\]
A positive value of $R^2_{\text{OOS}}$ indicates that the ARDL model achieves lower forecast error than the AR baseline, with the magnitude corresponding to the percentage improvement in forecast accuracy.

Finally, to formally test whether the augmented model significantly outperforms the baseline in a nested setting, we apply the Clark–West test. The test statistic is given by
\[
CW = \frac{\bar{f}}{\sqrt{\hat{V}(\bar{f})}},
\]
where $\bar{f}$ is the mean of the adjusted forecast accuracy differentials and $\hat{V}(\bar{f})$ is its heteroskedasticity- and autocorrelation-consistent (HAC) variance estimate. Statistical significance of the $CW$ test provides evidence that including discussion data yields incremental predictive gains.  

\end{document}